\documentclass[a4paper,11pt,twoside]{article}
\usepackage{amscd}
\usepackage{url}
\usepackage{fancyhdr,latexsym,amsmath,amsfonts,amssymb,amsbsy,amsthm,url}%
\usepackage{natbib}
\usepackage[hscale=0.7,centering]{geometry}
\usepackage{graphicx,epsfig, xy}
\usepackage{hyperref}
\usepackage{lscape,fancybox}
\usepackage{color}

\numberwithin{equation}{section}

\theoremstyle{remark}
\theoremstyle{definition}

\allowdisplaybreaks

\begin{document}
\def\shorttitle{COVID-19 endemicity}
\def\shortauthors{Bhattacharjee and Bose}
\title
{Modelling COVID-19-III: endemic spread in India}
\author{
\parbox[t]{0.60\textwidth}{Madhuchhanda Bhattacharjee\\
School of Mathematics and Statistics\\ University of Hyderabad\\ Gachibowli\\ Hyderabad 500046\\ India\\
email: mbsm@uohyd.ernet.in}
\hspace{.05\textwidth}
\parbox[t]{0.45\textwidth}{Arup Bose\footnote{Research supported by J.C. Bose National Fellowship, Department of Science and Technology, Government of India.} \\
Stat-Math Unit, Kolkata\\
Indian Statistical Institute\\
203 B.T. Road\\
Kolkata 700108\\ India\\
email: bosearu@gmail.com}
}
\date{\today}
%



\maketitle
\begin{abstract} \noindent
{\footnotesize 
A disease in a given population is termed \textit{endemic} when it exhibits a steady prevalence. We address the pertinent question as to what extent COVID-19 has turned endemic in India. 

There are several existing models for studying endemic behaviour, such as the extensions of the traditional temporal SIR model or the spatio-temporal endemic-epidemic model of Held et al. (2005) and its extensions. 

We propose a ``spatio-temporal Gravity model'' in a state of the art generalised linear model set up that can be deployed at various spatial resolutions. 
In absence of routine and quality covariates in the context of COVID-19 at finer spatial scales, we make use of extraneous covariates like air-traffic passenger count that enables us to capture the local mobility and social interactions effectively. This makes the proposed model different from the existing models. The proposed gravity model not only produces consistent estimators, but also outperforms the other models when applied to Indian COVID-19 data.

	}
\end{abstract}
\vskip10pt

\vskip10pt

\normalsize
\section{Introduction}\label{section:intro} 
COVID-19 hit the world in December 2019, and there has been a large volume of research articles devoted to its study from numerous angles. For a brief background on the history and spread of COVID-19 in India, the reader is referred to \cite{mbab2021}, and the references therein. Data on COVID-19 in India is primarily available in the form of three time series of  \textit{Cases} ($C$),  \textit{Recoveries} ($R$), and \textit{Deaths} ($D$). 

The most commonly used model for this data is some form of the temporal SIR model--with added compartments depending either on any additional information which may be available, or on the required policy prescriptions. When fitted dynamically, these models are capable of very good short term predictions. Most models on the spread of COVID-19 in India are built from the \textit{temporal} view point. 

India is often referred to as a sub-continent for its size and diverse nature of the regions. The progression of COVID-19 cases in the sub-continent over time has shown a wide variation across its 36 \textit{states and union territories}, henceforth collectively referred to as \textit{regions}. This is evident from the plots and discussions in \cite{bhattacharjee}, \cite{ranjan} and \cite{mbkdab2021}, as well as other sources. 

It is naturally expected that the propagation of COVID-19 to neighboring locations would lead to a spatial pattern. Areas which are connected, either geographically, or via movement of people by railways, road, air and  waterways, would influence each others' caseload. Moreover, due to many factors, including the natural topography of India, regions of India have varying degrees of accessibility and connectivity. As is well known, temporal SIR and related models cannot unearth the spatial dependence that may be present in a time series data.
 
On the other hand, it is widely accepted that taking spatial dependence into account can significantly impact statistical analysis of epidemics. See for example \cite{cliford}. \cite{kirby2017advances} provides an excellent review of spatio-temporal methods in epidemiological data. Incidentally, spatio-temporal analysis of US and European COVID-19 data have been made by several authors. See for instance \cite{sannipilla} and \cite{mollalo}. For Indian COVID-19 data, it is natural to consider \textit{spatio-temporal} models at the regional level, with \textit{districts} (sub-regions) being considered as the primary units in each region. 

To develop these models for India is a challenge for several reasons, not the least of which is the paucity and unreliability of data on the relevant \textit{covariates}, and on the auxiliary variables. The extent and accuracy of this information varies across the regions, and hence not every model is meaningfully implementable in every region. We are aware of a few attempts such as \cite{mbkdab2021} and \cite{ganesansubramani2020, ganesansubramani2021}.

When a disease within a given geographic region exhibits a steady prevalence in the population, then it is termed \textit{endemic}. An endemic behaviour is connected to it being seasonal, so that the disease incidence sees spikes around the same time every year--an obvious example is the common flu in different parts of the world. There is a strong opinion (see for instance \cite{phillips2021}) that COVID-19 will eventually turn endemic. A pertinent question is whether we are witnessing the onset of its endemic behaviour. Some researchers, for instance \cite{Bjornstad2020}, have estimated the time to reach an endemic equilibrium for COVID-19 as four years.

The traditional endemic models are based on modified versions of the SIR models. The endemic-epidemic (EE) models are more sophisticated spatio-temporal models. They are time series models for multivariate surveillance counts, and were introduced by \cite{heldohl2005}. Subsequently, these have been extended by several authors. For these models, the counts for the different variables are postulated to progress over time in a Markovian manner. While the parameters across the regions are in general different, some of them may turn out to be identical. However, the regional variations and apparent disharmonious occurrence of COVID-19 in different parts of India cause difficulties with the direct applications of such models to this sub-continent. 

COVID-19 has been observed to have a  strong epidemic component (see for example \cite{giulianidickson2020}), with noticeable spatial variation. A natural approach would be to employ a spatio-temporal model at a sub-regional level. This would allow for possible forecasts on various levels of spatial aggregation, namely sub-regional, regional and also perhaps national. Thus it would be worthwhile to derive a meaningful modelling framework at a finer geographic resolution that would provide consistent and meaningful inference on some of the essential aspects of this pandemic.

We have succeeded in developing spatio-temporal models for the Indian district level data within each region and have investigated their performance with two specific objectives. The first is to connect public mobility, social connectivity, and local behaviour with the extent of occurrence of COVID-19. The second objective is to assess how far it has reached an endemic state at the individual regional levels as well as collectively.

We consider the publicly available daily COVID-19 incidence data at the district level. Unfortunately, there are multiple issues with the available data, and in particular, it is affected by many factors that are extraneous to the epidemic. Thus we are forced to fall back on data on surrogate covariates that are affected by, as well as are influencing the epidemic. Data on these surrogates could possibly be collated by independent resources.  While theoretically there are many possibly useful covariates, in practice there is a stark lack of data on such  covariates at required space and time resolution. 

The air-traffic passenger data is publicly available, and we consider this as an important surrogate. 
It stands to reason that the observed spatial fluctuations in the counts for cases are often influenced by the local social events, such as festivals. At the same time, these very factors also influence the corresponding “air-traffic” covariate. Hence, in absence of any other reliable local spatial predictors, this surrogate covariate could play a significant role. However, air-traffic data is available only on a monthly basis, and there is a lag of a few weeks in its availability. To use this data meaningfully, we aggregated the COVID-19 incidence data at the monthly level.

For the second aspect, since the extent of endemicity is somewhat subjective, we use different models to assess this, and compare our estimates across the different novel models that we employ.

The remaining part of this article is organised as follows. In Section \ref{section:model_endemic} we review the available literature on spatio-temporal models 
for epidemiological data, especially those with endemic components. Section \ref{section:proposed_models} contains our proposed gravity models. Along with these novel models we also consider modifications of the existing models. In Section \ref{section:results}, we present the findings from these models, based of COVID-19 data from the Indian sub-continent, and demonstrate the superiority of their performance The final Section \ref{section:conclusions} contains discussion along with possible extensions of the current work.

\section{Modeling endemic behaviour}\label{section:model_endemic}

There have been the two major paths explored for modelling endemic patterns. The first of these include the mathematical and statistical models that extend the  well-known SIR model. We give an example of such an effort by  \cite{Bjornstad2020}. The second is the wholly probabilistic modelling pioneered by  \cite{heldohl2005}, which we shall discuss in details. 

\textbf{Extension of SIR models}

As mentioned earlier, the SIR models are among the most popular epidemic models. The original SIR models can also be extended by taking into account the number of \textit{exposed} individuals (see \cite{brauer2019}), say $E$. Since there is a latent period between being infected and becoming infectious, the trajectory of counts for $C$ will be in a lag with that for $E$ by this latency period.  With the reproduction number $R_{0} > 1$, both $E$ and $I$ are predicted to eventually stabilize at an \textit{endemic} equilibrium (see \cite{Bjornstad2020} for an illustration).

\textbf{Endemic-Epidemic (EE) models}

A study by \cite{colongonz} indicates that use of multiple data streams arising from surveillance activities can be a useful approach to disease detection. They opine that, syndromic surveillance complements traditional public health surveillance, by collecting and analysing health indicators in near real time. It is imperative that appropriate statistical techniques are used to analyse such data.

The endemic-epidemic (EE) model was introduced by \cite{heldohl2005}. These were subsequently extended 
in \cite{paulheldtoschke2008}, \cite{heldpaul2012}, \cite{meyerheld2014} and \cite{meyerheldhohle2017}. These models have also been effectively applied to several other epidemiological problems. See \cite{dunbarheld2020} for a review.

The EE model is based on the motivation that, while the epidemic component enables the capture of occasional outbreaks, the endemic component should explain the baseline rate of cases that is persistent with a stable temporal pattern (see \cite{adegboye2017}). Suppose we have weekly data on counts for different variables in different regions. Then, we assume the Markovian structure, so that  the counts for the $t$th week depend only on those for the $(t-1)$th week. This entails assuming implicitly that the time between appearance of the symptoms in successive generations is the same, and equals the observation interval of a week. However, in reality, this \textit{serial interval} may vary randomly across infection events and may be longer than a single observation interval (see \cite{becker2015}, p.156). Other factors can also introduce dependencies on time points $t-2, \ldots , t-D$ and $D$ can be quite high. This limitation was addressed by  \cite{bracherheld2020} who introduced flexible weighting schemes for past incidences. In particular, they suggested the \textit{shifted Poisson, triangular} and \textit{geometric weights} based on the different empirical distributions of the serial intervals.

Consider a specific region with $P$ districts. Let the number of cases in district $i$ for week $t$ be denoted by $Y_{i,t}$, $i = 1, \ldots , P$ and let $\textbf{Y}_{t}$ be the vector of these values. A simple example of an EE model is the following: first, conditionally on the past, $Y_{i,t}$ and $Y_{j,t}$ from different districts (units) $i$ and $j$ at time $t$ are assumed to be independent. Second, each $Y_{i,t}$,  conditionally on the past, is assumed to follow a negative binomial distribution. That is, 
\begin{equation}\label{eq:bracher-heldmodel}
Y_{i,t} | \textbf{Y}_{t-1}, \textbf{Y}_{t-2}, \ldots \sim  \text{NB}(\lambda_{it}, \psi_i), \ 1\leq i \leq P, \ 1\leq t \leq T,
\end{equation}
with conditional mean $\lambda_{it}$  and over dispersion parameter  $\psi_i$. The $\{\lambda_{it}\}$ is further modeled as 
\begin{equation}\label{eq:conditional_mean}
\lambda_{it} = \nu_{it} + \phi_{it}\sum_{j=1}^P \lfloor w_{ji} \rfloor\textbf{Y}_{j,t-1}, \ 1\leq i \leq P, \ 1\leq t \leq T.
\end{equation}
Here the parameter $\nu_{it}$ is the endemic component, and captures the number of infections that are not directly linked to the observed cases from the previous week. The remaining autoregressive term in (\ref{eq:conditional_mean}) is the epidemic component and describes how the incidence in district $i$ is linked to previous cases $Y_{j, t-1}$ in districts $j = 1, \ldots , P$.

The conditional variance is thus $\lambda_{it}+ \psi_{i} \lambda_{it}^2$ where $\{\psi_i\}$ are the over dispersion parameters. When  $\psi_i = 0$, the conditional variance reduces to $\lambda_{it}$. A common simplification is to assume that all the over dispersion parameters $\psi_i$ are identical, say, equal to $\psi$. 

The parameters $\{\nu_{it}\}$ and $\{\phi_{it}\}$ are constrained to be non-negative and modeled in a log-linear fashion, for instance with sine-cosine terms to account for seasonality (\cite{heldpaul2012}). Long-term temporal trends and covariates such as meteorological conditions (\cite{chengluwulihu2016}, \cite{bauerwakefield2018}) or vaccination coverage (\cite{herzogpaulheld2011}) could also be included. The coupling between the districts is achieved by the $P^2$ weights $w_{ji}$ which enter in (\ref{eq:conditional_mean}) after normalization as:
\begin{equation}\label{eq:wij_normalised} 
\lfloor w_{ji}\rfloor = \dfrac{w_{ji}}{\sum_{h=1}^P w_{jh}}.
\end{equation} 
Unrestricted estimation of all $P^2$ weights is usually unstable in practice, so more parsimonious and epidemiologically meaningful parameterizations have been introduced. Specifically, the weights can be based on social contact data for spread across age groups (\cite{meyerheld2017}), or on the geographical distance between cases to describe spatio-temporal spread (\cite{meyerheld2014}). In the latter case the weights can be specified through a power law 
\begin{equation}\label{eq:wij_powerlaw}
w_{ji} = (o_{ji} + 1)^{\rho},
\end{equation}
where $o_{ji}$ is the path distance between the regions $j$ and $i$ ($o_{ii} = 0$ for all $i$, $o_{ji} = 1$ when  $i$ and $j$ are direct neighbors, and so on), and $\rho$ is a decay parameter to be estimated from the data. The power law formulation is motivated by human movement behaviour (see \cite{brockmannhyfnagelgeisel2006}), and has been found to be an efficient way of capturing spatial dependence.

\textbf{Modified EE model}
 
The EE model of \cite{berlamannHaustein2020} is more advanced. 
It allows for possible over dispersion due to under-reporting or unobserved covariates that affect the disease incidence. As before, 
suppose that the conditional distribution of $Y_{i,t}$, given the past history up to  time $t-1$ which is symbolized as the $\sigma$-field $F_{t-1}$, is negative binomial with parameters $\mu_{i,t}$ and $\psi_{i}$. We write 
$$Y_{i,t}|F_{t-1}\sim NB(\mu_{i,t}, \psi_{i}),  \ 1\leq i \leq P, \ 1\leq t\leq T.$$

The parameter $\psi_{i} > 0$ is the \textit{over dispersion parameter}, so that the conditional variance of $Y_{i,t}$  is given by 
\begin{equation*} \text{Var}[Y_{i,t}|F_{t-1}]= \mu_{i,t} \left( 1 + \psi_{i}\mu_{i,t} \right).\end{equation*}
The \textit{mean} $\mu_{i,t} $ is given by 
\begin{equation}
\mu_{i,t} = \lambda_{i,t} \sum_{d=1}^{D} u_{d}Y_{i,t-d} + \phi_{i,t} \sum_{d=1}^{D} \sum_{j \neq i} u_{d} \omega_{j,i}Y_{j,t-d} + \nu_{t}, \ 1\leq i \leq P, \ 1\leq t\leq T.\label{eq:mean}
\end{equation}

The three summands on the right side of (\ref{eq:mean}) correspond to the \textit{epidemic-within districts}, the 
\textit{epidemic-between districts} and the \textit{endemic} components. 

The within component first summand is modeled as an autoregression of the number of cases in district $i$,  on a weighted sum of the past number of cases up to day $t-D$. The autoregression parameter is $\lambda_{i,t}$.

The between component second summand attempts to capture the spread of a disease across different locations. It is taken as a regression of the number of cases in district $i$ on a weighted sum of the past number of cases up to time $t-D$ in other districts $j \neq i$. The autoregressive parameter $\phi_{i,t}$ is district specific, and may depend on covariates. The component is completed by a system of weights that have two factors; while $\{u_d\}$ are the same as for the within component, the parameters $\{\omega_{j,i}\}$ account for the spatial distance between districts. 

The last term $\nu_t$ is the endemic component that describes the seasonality of the difference between regions, and typically it is modeled by capturing the seasonal variations of the disease through a harmonic wave: 
\begin{equation}
\log \nu^{END}_{t} = \alpha_0 + \eta t + \gamma \sin (\omega t) + \delta \cos (\omega t),
\end{equation}
with $\alpha_0$ being a constant, $\eta t$ being a time trend and $\gamma sin(\omega t) + \delta cos(\omega t)$ capturing possible seasonal variation of the endemic component as is typical for many viral diseases. The parameter $\alpha_0$ could also be equal to $\alpha_i$ so that it varies across districts. The parameter $w$ could then be $2\pi/365$ as Fourier frequencies corresponding to daily data.

The EE modelling approach has been used in \cite{grimeeheld2022}. 
More recently, \cite{celani_giudici_2021} applied this model to data from Italy, taking several covariates, such as the population density,
a Stringency Index, a Testing Policy index, and an indicator for weekend days. Two of their key covariates were available at only the national level.


\section{Proposed models}\label{section:proposed_models}

As mentioned earlier, the objectives here are twofold. The first is to be able to connect public mobility, social connectivity, and local behaviour with the extent of occurrence of COVID-19. In a noticeable contrast to theoretical possibilities of such candidate covariates, in reality there is a stark lack of data at required space and time resolution for the relevant covariates. 

The second objective is surveillance of this epidemic to assess whether it has reached a manageable state. Since the Indian sub-continent is extremely heterogeneous in its socio-demographic profile, this would be key in assessing the status of the epidemic at the individual regional levels as well as collectively.

As mentioned earlier we will model the COVID-19 incidence data using surrogate covariates like air-traffic passenger data which is publicly available and collated by independent resources. It stands to reason that the observed spatial fluctuations in the counts for cases are often influenced by the local social events, such as festivals. At the same time, these same factors also influence the corresponding “air-traffic” covariates. Hence, in absence of any other reliable local spatial predictors, this data could play a significant role. Unfortunately, the air-traffic data is available only on a monthly basis, and in addition there is a lag of a few weeks in its availability. To use this data gainfully, we aggregated the COVID-19 incidence data at the monthly level.

\textbf{Gravity model}

As is commonly done, we apply a negative binomial `observation-driven' model for count data. The proposed model for (log-)mean has effectively three components. The first component captures the purely time effects, the second captures the purely spatial effect, and the third is a combination of spatio-temporal effects reflecting the local variations.

We also borrow ideas from the gravity models. These models can be traced to \cite{ravenstein} and \cite{zipf1946p}, and they have become popular due to the subsequent work of \cite{tinbergen}. See \cite{anderson} for a review of these models. In a regression context, we may frame this model as follows: consider locations $i$ and  $j$ between which some form of transfer of population, say  $Y_{ij}$ occurs. Assume that information on the total movement from $i$ is available as $U_i$ and total arrival at $j$ is available as $V_j$. Additionally let $d_{ij}$ be the ``distance'' between the two locations. Then we postulate that
		\begin{equation}
			Y_{i,j}= \theta_0 U_i^{\theta_1} V_j^{\theta_2} {d_{ij}}^{\theta_3}\eta_{ij}, \label{eq:gravitymodel}
		\end{equation} 
where $\theta$'s are unknown parameters and $\eta_{ij}$ is an error term. Since  $Y_{ij}$ are non-negative, a modified version of the above model stipulates that the  mean $\mu_{ij}$ of $Y_{ij}$ is given by:
		\begin{equation}
			\log \mu_{ij}= \theta_0  + {\theta_1} \times U_i  + {\theta_2} \times V_j + {\theta_3} \times {d_{ij}}  \label{eq:gravitymodelmean}.
		\end{equation} 
Even though the above model assumes a power decay $d_{ij}^{\theta_3}$ between the districts $i$ and $j$, more generally, it may be modeled by a decreasing \textit{distance decay function} $f$. Some specific choices of $f$ that have been used in the literature are the (i) \textit{power} decay $f(d_{ij})=d_{ij}^{-\alpha }$, (ii) \textit{exponential-normal} decay $f(d_{ij})=e^{-\alpha d_{ij}^2}$, (iii) \textit{exponential-square-root} decay  $f(d_{ij})=e^{-\alpha \sqrt{d_{ij}}}$, and (iv) \textit{exponential} decay $f(d_{ij})=e^{-\alpha d_{ij}}.$ For other choice of distances, see \cite{berlamannHaustein2020}. 


For count data regression when an offset variable is used, then the corresponding regression coefficient is prefixed to be 1. Such a model then represents rates in comparison to counts.  In the literature it has been suggested to use the population size as an \textit{offset term} (see \cite{cox1981}). 
However due to various shortcomings in the data available, we do not use the population size as an offset, and instead assign a set of geographic region specific (e.g. district level, regional level) parameters, which would effectively serve the same purpose. An endemic component is also included in the model as suggested earlier. 

We employ the gravity models in two variations. Suppose there are $K_i$ airports within the $i$-th region. In variation 1, at the $t$-th time point, for the $j$-th district of the $i$-th region, the log-mean of incidence is given by 
\begin{equation}\label{eq:gravity1}
\lambda_{ijt} = \nu_{it} + \phi_{ij;} + \sum_{k=1}^{K_{i}} \theta_{ik} D_{ijk}*X_{ikt}, \ 1\leq j \leq R_{i}, \ 1\leq i \leq P, \ 1\leq t \leq T,
\end{equation}
where $D$ contains the distances with exponential decay, and $X$ contains the air-passenger-traffic data. 

In variation 2,  we have a composite gravity model, 
where,
\begin{equation}\label{eq:gravity2}
\lambda_{ijt} = \nu_{t} + \phi_{ij} + \sum_{k=1}^{K_{i}} \theta_{ik} D_{ijk}*X_{ikt}, \ 1\leq j \leq R_{i}, \ 1\leq i \leq P, \ 1\leq t \leq T,
\end{equation}
and $D$ and $X$ are as above. 
 
Both models (\ref{eq:gravity1}) and (\ref{eq:gravity2}) can also be implemented with an intercept term, while adjusting the $\nu_t$ and $\phi_{ij}$ parameters accordingly.

\textbf{Endemic-epidemic type models }

For local COVID-19 infection counts, following the ideas of \cite{heldohl2005}, we develop a class of spatio-temporal regression models that have both endemic and epidemic components.

The scarcity of useful covariates led us to use air-traffic data for the Gravity models described earlier, and this data is available only at a monthly frequency. Note however that the EE-type models are (spatial-)autoregressive in nature. Since COVID-19 is a rapidly evolving epidemic, such an autoregressive model at the monthly time scale is not expected to capture its evolution. Therefore the relevant parameters in these models are then estimated using the daily occurrence data from various regions.


In our models, the expected conditional mean spread of the infectious disease is decomposed into endemic and epidemic components. However, deviating from the existing methods, we implemented these models at various levels of aggregation of the parameters. 

For example, for each individual region $i$, the expected conditional mean $\mu_{it}$ at a time $t$ for that region is postulated as:
$$\log \mu_{it} = \nu^{EPI}_{it} + \nu^{END}_{it},$$
with an over-dispersion parameter $\psi > 0$. 

In contrast, in a separate model, we assume that the endemic component $\nu^{END}_{it}$ at the time $t$, is common to all the regions, regardless of the infection history of district $i$ and its neighbors. Let the common value be $\nu^{END}_{t}$. In that case $\log \mu_{it}$ reduces to 
$$
\log \mu_{it} = \nu^{EPI}_{it} + \nu^{END}_{t}.
$$
Interestingly \cite{berlamannHaustein2020} further multiply $\nu^{END}_{t}$ by the size of the local population. While this would be a preferred structure, shortcomings in the data prevents us from using this modification.

As described earlier, most commonly used structure for the epidemic part of the regression are primarily autoregressive, where the spatial influences are also modeled as a lagged regression (see \cite{heldohl2005}, \cite{bracherheld2020}). This enables the capture of path dependencies and of self-exciting behaviour that are known to be common with infectious diseases. Thus for the epidemic part of our model, we adhere to autoregressive components similar to those proposed in \cite{heldohl2005}. For the spatial autoregression component, we use the entries of the spatial lag-$1$ adjacency matrix as weights.

Depending on the context, additional covariates such as age and gender-specific effects, contagion in nearby districts in the geographic and social space, as well as latent heterogeneity between the districts are taken into account. See \cite{fritzkauermann2020}) for an example. They also provide an extensive covariate based model for the endemic component.
In our models we have used region specific parameters. 

\cite{berlamannHaustein2020} use a region specific random effects component in the epidemic part of their model, in conjunction with the spatial autoregressive part. In contrast we have employed a fixed effect parameter for the effects of regions, and the varying population sizes of the individual regions are subsumed in this parameter.

\section{Results}\label{section:results}


The data sources used for implementation of the models are \textit{https://data.incovid19.org} for COVID-19 incidence data, and \textit{https://www.aai.aero} for air traffic passenger data. We have compared various alternative model specifications on these data. For model comparison and model fit to assessment, we have used both, the Akaike Information Criterion (AIC) and the Nagelkerke pseudo-$R^2$. Due to widely different nature of the models implemented in this data, we have presented the pseudo-$R^2$ values. The fact that these values are naturally limited to a common range of values, eases the comparison across various models. It is to be noted that the gravity models have been implemented with monthly aggregated data whereas the spatial-autoregressive-type EE model uses daily incidence data.

The main commonality of the proposed Gravity and EE-type models is that they all have an endemic component, at regional or national level, denoted by $\nu_{.}$. The remaining components of these models pertain to capturing the epidemic behaviour. We have applied various alternative specifications of the epidemic components of the model, which serve to check the robustness of our conclusions with regard to the estimation of the endemic pattern. 

To assess the presence of a spatially-consistent endemic component, we implemented the gravity model at the regional level with individual region specific endemic parameters $\{\nu_{it}\}$, as well as an overall model based on all regions with  shared endemic parameters $\{\nu_t\}$. We compared our findings with the competing EE-type model. Further, to assure us against spuriousness, we also used multiple data sources and checked the consistency of the endemic-parameter estimates, in terms of  both, the pattern as well as scale of the estimated values.

\begin{figure}[h!]
\vskip-20pt
\begin{center}\includegraphics[height=70mm, width=70mm]{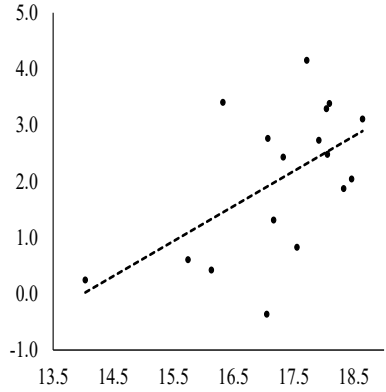}
\end{center}\vskip-25pt
\caption{\small{Gravity model for individual regions: log population (in horizontal axis) against estimated location parameter (in vertical axis).}}
\label{fig:Log_Pop_Intrcpt}
\end{figure}

\textbf{Gravity model: offset parameter estimates}. 
As mentioned earlier, we did not include the (log)-population in the model, and instead used a separate parameter to capture the location. We have validated our choice of structure for the various parts of our model (see expression (\ref{eq:gravity2})). For example, based on the proposed model (\ref{eq:gravity1}),  for each region, we obtained estimates for locations as sum of the estimated intercept parameter(s), and the average of district level parameters from the respective region. From Figure \ref{fig:Log_Pop_Intrcpt} we observe that, while there is a reasonable relationship between these estimated location parameters and the corresponding log-population values, there are some departures. This justifies our choice of the model structure in this respect. We have also assessed the relationship with the population density, and the observed pattern is similar to that in Figure \ref{fig:Log_Pop_Intrcpt}.

\begin{figure}[h!]
\vskip-5pt
\begin{center}\includegraphics[height=100mm, width=100mm]{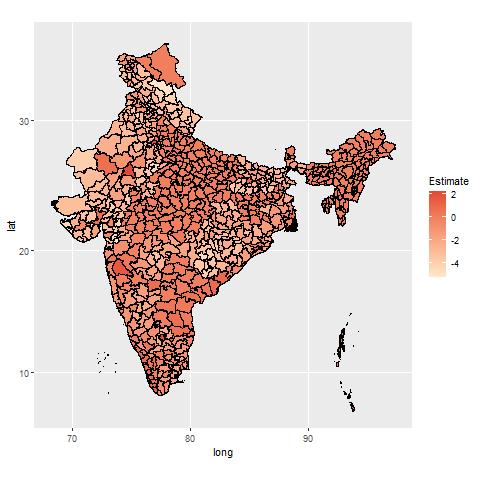}
\end{center}
\vskip-25pt\caption{\small{Gravity model at the regional level: estimated district level parameters for the 17 regions. For the districts from other regions, the overall average has been used.}}
\label{fig:Map_dst_params}
\end{figure}

\textbf{Gravity model: spatial parameter estimates}. 
In the epidemic part of the proposed model, the very first components are the spatial/regional effect parameters, $\phi$'s. In Figure \ref{fig:Map_dst_params} we have presented the estimated parameters for various districts in India under the Gravity model implemented for the 17 regions where districts level COVID-19 data is available, in a combined manner. The interpretation of the parameters (and their estimates) are on the log-scale.

\begin{figure}[h!]
\vskip-25pt
\begin{center}\includegraphics[height=120mm, width=150mm]{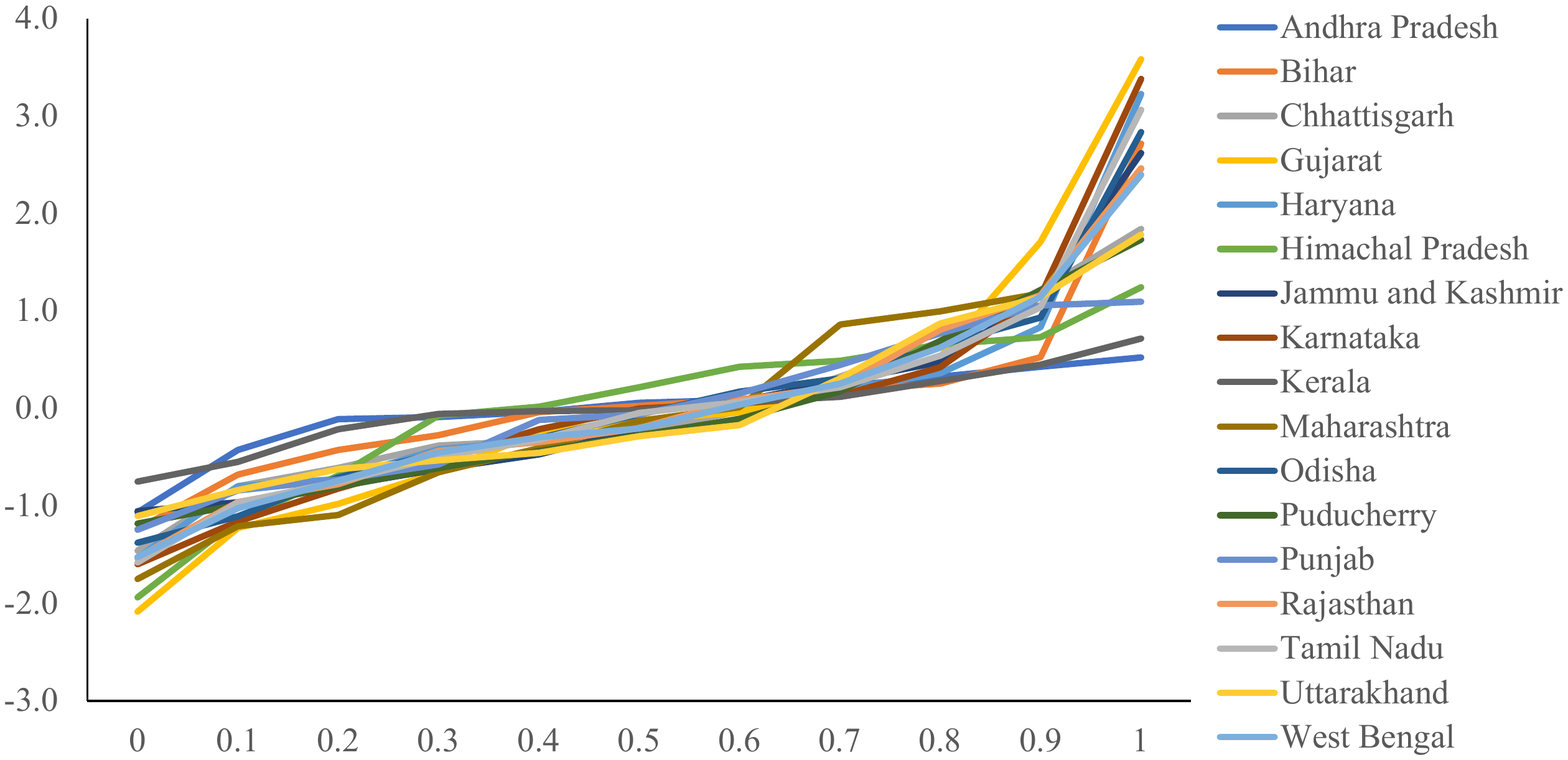}
\end{center}
\vskip-30pt\caption{\small{Gravity model at the regional level: estimates of quantiles of the location adjusted district parameters for each of the 17 regions.}}
\label{fig:Qntile_dst}
\end{figure}

\textbf{Gravity model: distributional commonality of the spatial parameters}. 
For each region, once we take away the location effect as described above, the centered estimates of the district level parameters appear to have similar distribution. In Figure \ref{fig:Qntile_dst}  we have provided the estimated quantiles of the centered district level parameters for each region. This demonstrates the commonality of the underlying behaviour of COVID-19 occurrences across regions.

\textbf{Gravity model: robustness of spatial pattern against implementation level}. 
These models are flexible enough to be implementable either in an individual region or as a combined model over multiple regions. The distributional behaviour of the district level estimated parameters appear to be extremely similar in nature for the individual and the combined models. These results are not shown to avoid repetition. Since the findings appear to be not sensitive to the level of implementation, it provides the assurance of the robustness of the pattern captured by the Gravity model.

\textbf{Gravity model: effect of gravity components in model performance}. 
As described earlier, the gravity component is expected to capture the difference in the local behaviour of the social and mobility aspects of the populations. Air-traffic has been considered as a proxy variable, since data on other possible covariates are not available publicly/systematically for modeling purposes. One may raise the issue that since life has returned to near-normalcy, whether these variables can still be useful for modeling COVID-19.  In Table \ref{table:1}, we have summarised the performance of a selected set of models. We can safely conclude from these values that, these variables definitely contribute significantly in the various model combinations applied on this data.
\begin{table}[h!]
\begin{center}
\caption{\small{Effect of gravity components in model fit, assessed by pseudo-$R^2$.}}\label{table:1} 
\vskip3pt
\begin{tabular}{lcc} \hline
& \multicolumn{2}{c}{Gravity component} \\ \cline{2-3}
Other covariates in the model & Without &  With \\ \hline
Regional level effects                                          & 0.182 & 0.290 \\
District level effects                                       & 0.363 & 0.497 \\
District level effects and seasonal components & 0.457 & 0.556  \\
District level effects and endemic components  & 0.826 & 0.997  
\end{tabular} 
\end{center}
 \end{table}

\textbf{Gravity model: gravity component with complete connection of sources and destinations}. 
Another relevant aspect in modeling the epidemic component is whether it was possible to travel across the regional boundaries to a neighboring region for air travel. We implemented a separate model which allows for the inter-regional travel, leading to a gravity model with completely connected source-destination network dampened only by distance decay. We found the model estimates to be similar but the overall fit was poorer. This is not surprising since for a considerable period of the pandemic, various travel restrictions in each region had made cross-region travel to airports quite infeasible for most people.

\textbf{Gravity model: robustness of estimates against data source}. The availability of COVID-19 data for India has been volunteer driven. Thus there could be skepticism whether our conclusions are data source specific. We have tried our models on the popular/regular sources of COVID-19 data. We also extracted and cleaned data from the MS-Bing website \textit{https://www.bing.com/covid/local/india}. Irrespective of the data source, the estimated parameters and patterns have been extremely similar. Unfortunately since October 2021, data in the MS-Bing resource  has become erratic, and hence full comparison was not possible. The partial results have not been shown here.

\begin{figure}[h!]
\vskip-20pt
\begin{center}\includegraphics[height=90mm, width=110mm]{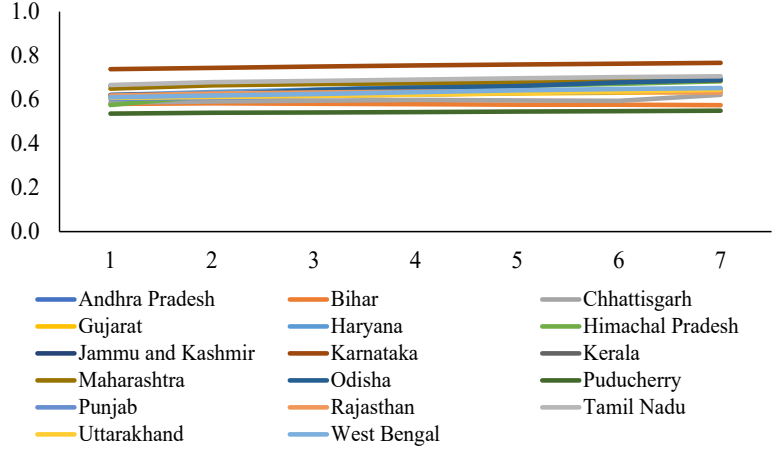}
\end{center}\vskip-25pt
\caption{\small{Nagelkerke pseudo-$R^2$ for regional endemic-epidemic type models with daily count data, for different autoregression lag choices.}}
\label{fig:EE_st_R2_lags}
\end{figure}

\textbf{EE model: lag structure of regional level history}. 
Before we arrived at an estimate of $\nu_{it}$ or  $\nu_t$ based on the EE model proposed by \cite{celani_giudici_2021}, it was essential to assess the lag effects of the time series of regional COVID-19 counts on each other. For this purpose, we used the daily COVID-19 occurrence data from each region, and implemented various lag structures for the autoregressive and the spatial effects parts of the models. We measured the effects of such changes by estimating the pseudo-$R^2$ for each of these models. Figure \ref{fig:EE_st_R2_lags} shows that there is little impact of higher order lags on the EE model, when measured in terms of the relative gain in Nagelkerke pseudo-$R^2$. Thus for subsequent implementations, we used a lag of order 1 for both local and spatial autoregression. 

\textbf{Endemic parameters: EE model with multi-region common parameters}. 
In the same EE model framework, instead of implementing the model for each region with their own endemic parameters $\{\nu_{it}\}$, we implemented a combined model with common endemic parameters $\{\nu_t$\}.

\begin{figure}[h!]
\vskip-10pt
\begin{center}\includegraphics[height=120mm, width=150mm]{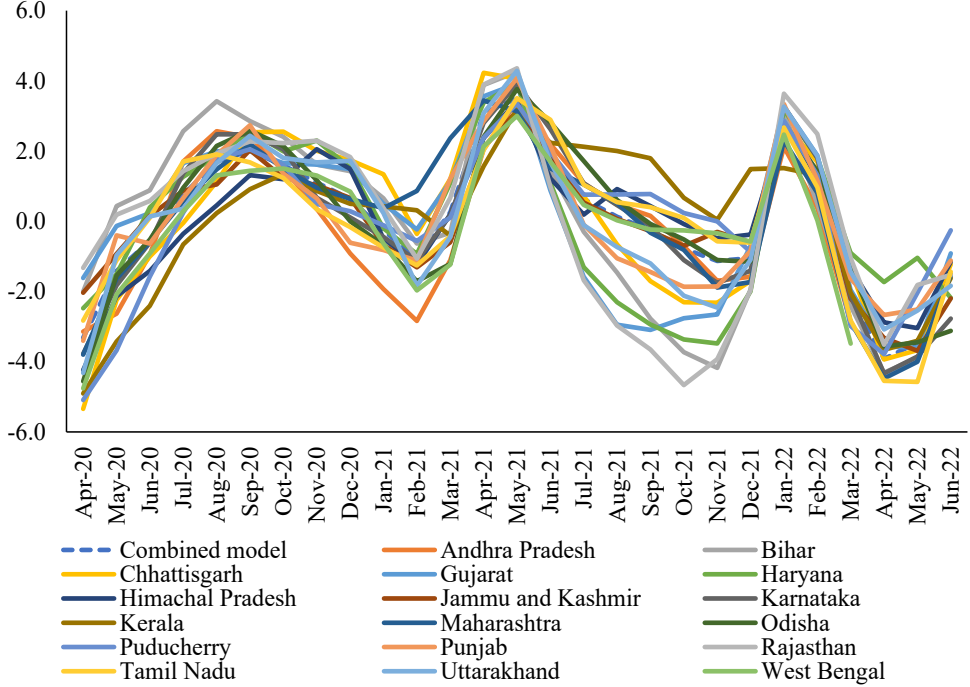}
\end{center}\caption{\small{Endemic parameter estimates under Gravity model for individual regions (labeled by region names), 
and all regions combined (labeled as 'Combined model').}}
\label{fig:Tp_St_Combo}
\end{figure}

\begin{figure}[h!]
\vskip-10pt
\begin{center}\includegraphics[height=100mm, width=120mm]{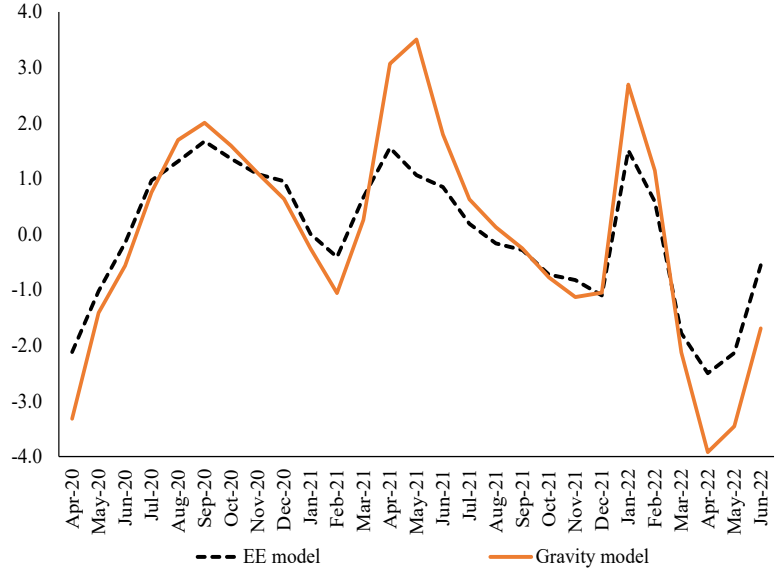}
\end{center}\caption{\small{Endemic parameter estimates under EE model and Gravity model based on multi region combined data.}}
\label{fig:Tp_EE_Combo}
\end{figure}

\textbf{Endemic parameters: gravity model with robustness against implementation level}. 
Figure \ref{fig:Tp_St_Combo} shows the estimated $\{\nu_{it}\}$ parameters, based on Gravity models applied to each region individually, and also the estimated $\{\nu_t\}$ parameters under the combined model. We observe that these parameter estimates show a strongly similar pattern over time, irrespective of the wide difference in the behaviour of the pandemic in different regions of India.
Similar comparison between the common $\{\nu_t\}$ parameters estimated under the two models, namely the EE model of \cite{heldohl2005} (see \cite{adegboye2017}) and the Gravity models, are quite similar in nature (see Figure \ref{fig:Tp_EE_Combo}).

\textbf{Endemic parameters: consistency under different model choices}. 
In Figure \ref{fig:Tp_EE_Combo} we present a comparative plot of the estimated endemic parameters, assumed to be common across the regions of India, estimated using the EE model as well as the proposed model (\ref{eq:gravity2}). It is to be noted that the EE model has been implemented on  the daily series, whereas the Gravity model utilized the monthly data. Thus the unit of time $t$ for $\nu_t$ are different for the two implementations. After a thorough comparison, it was felt that the mid-month estimates of $\nu_t$ based on EE-model applied to daily data is best for comparison with the estimates of the same from the Gravity model applied to monthly data. The estimates from the EE model and the Gravity model are shown in Figure \ref{fig:Tp_EE_Combo}, and it can be seen that they concur and show noticeable similarities. Thus, the estimates of the endemic parameters are persistently similar under different models/implementation and data sources. This enhances our confidence in the model and the estimation procedure.

\begin{figure}[h!]
\vskip-10pt
\begin{center}\includegraphics[height=90mm, width=110mm]{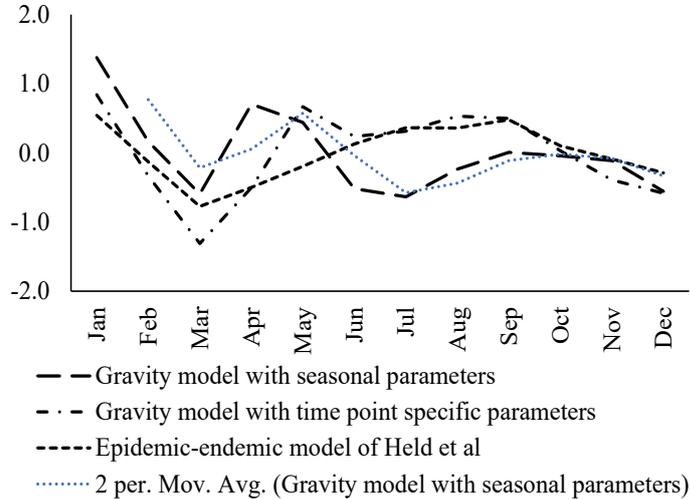}
\end{center}\caption{\small{Seasonal endemic parameter estimates under various models.}}
\label{fig:Tp_Ssnl}
\end{figure}

\textbf{Endemic parameters: assessment of seasonality}. 
One of the natural indicators of whether the epidemic has reached an endemic phase would be a strong seasonal component. To assess this, we modified the  models (\ref{eq:gravity1}) and (\ref{eq:gravity2}), where instead of a time point specific $\nu_t$ parameter, we implement a month specific set of parameters. If the endemic phase has indeed been reached, then this modified model would be able to achieve the same modeling precision with lesser number of parameters. In Figure \ref{fig:Tp_Ssnl} we have presented the estimated effects of months on COVID-19 log-mean counts, based on (i) the Gravity model for monthly data with seasonal parameters, (ii) the Gravity model based on monthly data with time point specific parameters, and (iii) the EE model based on daily data with daily parameters. We also applied a moving average of order 2 on the seasonal estimates, which provides a reasonable fit. The estimates of months from the Gravity and EE models are arrived at by combining the estimates from the respective months. From this figure we notice a consistency of pattern under the three modeling frameworks.

The pseudo-$R^2$ values are listed in Table \ref{table:2}. These values show that while all three modeling attempts arrive at a similar pattern, the seasonal model performs far worse than the models (\ref{eq:gravity1}) and (\ref{eq:gravity2}). This indicates that reduction in the number of parameters within the endemic part of the model is not yet possible. In other words, the endemic phase of COVID-19 has not been reached in India yet.
\begin{table}[h!]
\begin{center}
\caption{\small{Model fit assessment for selected models using Nagelkerke pseudo-$R^2$.}}
\label{table:2}
\vskip3pt
\begin{tabular}{ccc} \hline
\multicolumn{2}{c}{Gravity model} & EE model\\ \cline{1-2}
Seasonal & Non-seasonal & Non-seasonal \\ \hline
0.556 & 0.997 & 0.878
\end{tabular} 
\end{center}
 \end{table}

\section{Conclusions}\label{section:conclusions} 

The contributions of the Gravity type models proposed here are twofold. First, we used a novel set of covariates 
that account for temporal dynamic, latent effects and other covariates.  This enhances our understanding of the spread of COVID-19 at  a local level. Second, we have built this model within a state-of-the-art regression framework that allows ease of implementation for other applications. 

The extraneous covariates yielded an unexpected utility. Questions about the nature of collected and reported COVID-19 data have persisted over time. Many news articles have questioned what appeared to be the systematic under-reporting of COVID-19 cases from certain regions. Since the models presented here use externally collected covariates, these departures/lack-of-compatibilities within the data from temporal and/or spatial perspective  become clearly visible. A detailed reporting of such findings are beyond the scope of this article and hence have been omitted.
 
The underlying spatial pattern of COVID-19 is expected to be reasonably smooth, although local mobility and social behaviour can strongly influence the outcome. For most of the regions, this hypothesis is supported by plots of the estimated coefficients for the district level effects on the map. There is an evident degree of similarity in the underlying distributions of the district level parameters. Additionally, these district level parameters appear to be spatially smooth. In this regard the parameterization in our model overlaps with that proposed by \cite{alipour2021}, although their focus was to estimate the effect of ``working from home''.

This Gravity model is in spirit similar to the surveillance model introduced by \cite{heldohl2005}. Their extension of the generalized linear models to analyze surveillance data from epidemic outbreaks was further expanded to handle multivariate surveillance data by \cite{paulheldtoschke2008}. Useful modifications to account for seasonality and spatial heterogeneity was proposed in \cite{heldpaul2012}, and neighborhood information from social contact data was included in \cite{meyerheld2017}. It is agreed that these models have strong mathematical foundation. 

However, the proposed Gravity type model has novel explanatory features in addition to being mathematically rigorous. Our gravity model uses a part of the socio-economic behaviour in explaining the occurrence of cases, instead of using the evolution of disease process solely. This would be critical in situations 
where disease data collection mechanism is erratic, as is the case for India. In the end, the gravity component of the model successfully captures the local fluctuations, and the final fit to the data is extremely well. \cite{berlamannHaustein2020} had concluded that their disaggregated data showed that there is quite some variety in the relative importance of the endemic, the autoregressive epidemic, and the spatial endemic components. In our case, the endemic component is seen to be systematically similar across various regions of India, and the main variations are observed to come from local sources. 

Further modifications to our models could be explored but would require additional investigation. In one of our earlier works, we had employed novel clustering algorithms that explored and captured similarities between districts that are not necessarily spatially adjacent. Some possible explanations of such behaviour would be the shared latent non-spatial variables like climate, transport, and local culture. This could have been exploited in our current model too. In the literature, for example in \cite{dursogio2019} and \cite{dursovitale2020}, such techniques have been used on different proximity dimensions, such as the geographical and social space, to identify similar districts while taking into account spatial dependencies. Spatial dependencies have also been directly incorporated in the correlation structure in the context of spatial econometric models. See for example \cite{lesagepace2009}. This could also be a possible avenue to explore in extending these models.


\bibliography{spatio_covid_ref}
\bibliographystyle{abbrvnat}

\end{document}